\documentclass[10pt,aps,prd,showpacs,nofootinbib,twocolumn,superscriptaddress]{revtex4-2}

\usepackage{amssymb,amsmath,amsfonts}
\usepackage[usenames]{xcolor}
\usepackage[pdftex]{graphicx}
\usepackage[english]{babel}
\usepackage{amsmath,amssymb}
\usepackage{subcaption}  
\usepackage{multirow}
\usepackage{txfonts}
\usepackage{ulem}
\usepackage{ragged2e}
\usepackage[percent]{overpic}
\usepackage{mathtools}
\usepackage{comment}

\xdefinecolor{mylinkcolor}{rgb}{0,0,0.5}
\usepackage[
    bookmarksnumbered, bookmarksopen, bookmarksopenlevel=2,
    breaklinks=true,
    colorlinks=true, filecolor=mylinkcolor, citecolor=mylinkcolor,
    linkcolor=mylinkcolor, urlcolor=mylinkcolor, menucolor=mylinkcolor,
]{hyperref}
\usepackage{cleveref}

\usepackage{lipsum}

\begin{document}

\title{On the phase transition mechanism of spontaneous scalarization}

\author{Jo\~ao Vitor M. Muniz}
\email{joaomotta@id.uff.br}
\affiliation{Instituto de F\'isica, Universidade Federal Fluminense, Niter\'oi, RJ, 24210-346, Brazil.}

\author{N\'estor Ortiz}
\email{nestor.ortiz@nucleares.unam.mx}
\affiliation{Instituto de Ciencias Nucleares, Universidad Nacional Aut\'onoma de M\'exico, Circuito Exterior C.U., A.P. 70-543, M\'exico D.F. 04510, M\'exico}

\author{Raissa F.\ P.\ Mendes}
\email{rfpmendes@id.uff.br}
\affiliation{Instituto de F\'isica, Universidade Federal Fluminense, Niter\'oi, RJ, 24210-346, Brazil.}
\affiliation{CBPF - Centro Brasileiro de Pesquisas F\'isicas, 22290-180, Rio de Janeiro, RJ, Brazil.}

\date{\today}

\begin{abstract}
In certain modified gravity theories that include additional scalar degrees of freedom, compact objects such as black holes and neutron stars may undergo a process known as spontaneous scalarization, in which the scalar field is suddenly activated beyond a certain critical point. 
Since its discovery, it has been clear that this effect can be understood in many cases as a continuous phase transition, well described by the phenomenological Landau model. Recently, it has been pointed out that spontaneous scalarization can also manifest as a first-order phase transition. In this paper, we take a closer look at the nature of spontaneous scalarization as a phase transition, analyzing in detail cases where it occurs as either a second- or first-order transition, as well as a more unconventional scenario characterized by a negative scalar susceptibility. Critical exponents are explicitly computed, and implications for dynamical scalarization are discussed. Moreover, the dynamics of a first-order phase transition is probed through fully nonlinear numerical simulations.
\end{abstract}


\maketitle

\section{Introduction}
\label{sec:intro}

As electromagnetic and gravitational-wave observations expand our ability to study compact objects such as black holes and neutron stars, considerable effort has been devoted to testing general relativity and its alternatives in the strong-field regime that characterizes their vicinities \cite{Berti:2015itd}. A paradigmatic phenomenon in this context is the so-called scalarization effect, in which the scalar field present in a modified theory of gravity is activated only once a critical point is crossed (see Ref.~\cite{Doneva:2022ewd} for a review, and references therein).
Scalarization occurs for isolated compact objects when a critical configuration is reached within a sequence of equilibrium solutions. In binary systems, this effect can be enhanced, a phenomenon known as dynamical scalarization \cite{Barausse:2012da,Shibata:2013pra,Palenzuela:2013hsa}.

Shortly after its discovery in \cite{Damour:1993hw}, it became clear that, at least in certain theories and for certain ranges of theory parameters, scalarization can be understood as a continuous phase transition well described by the phenomenological Landau model (see, e.g., Sec.~II in \cite{Damour:1996ke}). Indeed, behind the name ``spontaneous scalarization'' was the close correspondence between this effect and spontaneous magnetization in ferromagnets \cite{Damour:1996ke}. This view has been fruitfully adopted in the development of effective models for dynamical scalarization, which explicitly use a Landau ansatz for the free energy near the critical point to construct an effective point-particle action describing the binary components \cite{Sennett:2017lcx,Khalil:2019wyy,Khalil:2022sii}.

However, as the set of theories and models exhibiting the characteristic features of spontaneous scalarization continues to expand, it is worthwhile to reassess its nature as a phase transition and examine whether---and under what conditions---it can be understood as a continuous phase transition well described by the phenomenological Landau model. In this work we aim to contribute to the broader effort of clarifying this interpretation.

We begin by framing the problem of spontaneous scalarization within the framework of Landau's phenomenological theory, analyzing in detail a ``typical'' case where scalarization is accurately described as a continuous phase transition. In particular, we confirm that the critical exponents align with the standard predictions of the Landau theory, and that a Landau ansatz provides a good fit for a suitably defined free energy near the critical point. 

We then examine scenarios where the Landau picture must be adjusted or even replaced. Specifically, we show that spontaneous scalarization can, in some cases, manifest itself as a first-order phase transition rather than a second-order (continuous) one. Examples of this type were recently discussed in Ref.~\cite{Unluturk:2025zie} for massive scalar-tensor theories, where it was argued that a first-order phase transition around the low-mass critical point may be a more typical situation than a second-order one, at least for a certain class of models.
We extend this interpretation to the phase transition around the high-mass critical point. 
By numerically constructing the Landau free energy, we firmly establish the interpretation of a first-order phase transition, further strengthening the analogy with material phase transitions \cite{Kuan:2022oxs}. 
Additionally, we carry out fully nonlinear numerical simulations that reveal both scalarization and descalarization processes within the first-order regime.

Finally, we explore a more unconventional scenario characterized by a negative scalar susceptibility. In this case, the commonly used free energy no longer meets the criteria of Landau's model. A negative scalar susceptibility also has intriguing implications for dynamical scalarization, which we discuss in some detail.

The paper is structured as follows. In Sec.~\ref{sec:pt} we review essential background material on phase transition theory. In Sec.~\ref{sec:NSinSTTs}, we define the classes of scalar-tensor theories under consideration and outline the construction of static neutron star solutions within these models. Section \ref{sec:SSPT} revisits the basis of the interpretation of spontaneous scalarization as a phase transition, reassessing the conventional case in which the transition is of second order. In Sec.~\ref{sec:first} we examine cases where scalarization manifests as a first-order phase transition, explicitly constructing the Landau free energy and investigating the associated dynamics through fully nonlinear numerical simulations. Section~\ref{sec:other} explores a more unconventional scenario, marked by a negative scalar susceptibility that challenges the applicability of the previously employed Landau free energy construction.
Section \ref{sec:conclusions} summarizes our main conclusions. We use natural units in which $G = c = 1$ unless stated otherwise.

\section{Basic elements of phase transitions}
\label{sec:pt}

The theory of phase transitions has evolved significantly over time, beginning with phenomenological models like Landau's mean-field theory, which provided a macroscopic description based on symmetry-breaking arguments. Later, the development of renormalization group techniques revolutionized the field by systematically addressing scale dependence and explaining the emergence of universality classes \cite{Goldenfeld:1992qy}. In this section, we provide a brief overview of Landau's theory, while leaving the adaptation of more advanced statistical physics tools to the problem of scalarization for future studies.

Landau's theory serves as a meta model for studying continuous phase transitions. It begins by identifying an order parameter, $\eta$, in the system under consideration, such that $\eta = 0$ for the high-symmetry (or disordered) phase and $\eta \neq 0$ for the low-symmetry (or ordered) phase. It also identifies a control parameter $\tau$ such that $\tau > 0$ (resp. $\tau <0$) in the disordered (resp. ordered) phase, and $\tau = 0$ at the critical point for the phase transition.
Landau's theory then assumes the existence of a (coarse-grained) free energy density $\mathcal{L}$, which depends on the order parameter (and other internal and/or external parameters) and is such that the state of the system corresponds to the global minimum of $\mathcal{L}$ with respect to $\eta$.  
Near the critical point, Landau's free energy is assumed to be an analytic function of $\eta$, admitting a power series expansion of the form:
\begin{equation}
    \mathcal{L} = a_{0} + a_{1} \eta + a_{2} \eta^{2} + a_{3} \eta^{3} + a_{4} \eta^{4} + \dots - h \eta \, ,
\end{equation}
where $h$ is a possible external field. The expansion coefficients are such that\;
    \begin{itemize}
        \item $a_{0} = \mathcal{L}[\eta = 0]$, i.e., the free energy value in the disordered phase ($\tau > 0$). This constant is generally  unimportant and can be shifted away.
        \item $a_{1} = 0$ since $\eta = 0$ is assumed to minimize $\mathcal{L}$ in the disordered phase for zero external field, i.e., $\partial \mathcal{L}/\partial \eta |_{\eta = 0, h = 0} = 0$.
        \item $a_{2} > 0$ for $\tau > 0$ and $a_2 < 0$ for $\tau < 0$, so that, near the critical point, $a_2 = a_2^{(1)} \tau + O(\tau^2)$.
        \item $a_{3}$ is often assumed to vanish from symmetry considerations; if non-zero, a first-order phase transition may take place.
        \item $a_{4} > 0$, otherwise the model, truncated at fourth order, would be minimized for $|\eta| \to \infty$. An extension of the model to higher orders will be discussed below, where this condition will be dropped.
    \end{itemize}
Taking the points above into consideration, a minimal ansatz for the Landau free energy density around the critical point is
\begin{equation}\label{eq:Landau_ansatz}
    \mathcal{L} = a_{2} \eta^{2}  + a_{4} \eta^{4} - h \eta \, .
\end{equation}

Around the critical point, thermodynamic quantities (i.e., derivatives of the free energy) exhibit a power-law behavior with respect to the control parameter. The exponents of these power laws are called critical exponents, which can be shared by different physical systems within the same universality class. The order parameter itself behaves as $\eta \propto |\tau|^{\beta}$ in the ordered phase; the ``heat capacity'' $C = - \tau \partial^2 \mathcal{L}/\partial \tau^2$ behaves as $|\tau|^{-\alpha}$, the ``susceptibility'' $\chi = \partial \eta/ \partial h$ behaves as $\chi \propto |\tau|^{-\gamma}$, and we have $h \propto \eta^\delta $ at the critical point ($\tau = 0$).

The Landau model predicts $\alpha = 0$ ($C$ displays a discontinuity at the critical point), $\beta = 1/2$, $\gamma = 1$, and $\delta = 3$. 
These values can be derived straightforwardly. For instance, minimizing the Landau free energy $\mathcal{L}[h=0]$ yields, in the ordered phase, $\eta = \sqrt{-a_2(\tau)/[2a_4(\tau)]}$. Considering the leading-order behavior of these coefficients, given by $a_2 = a_2^{(1)} \tau + O(\tau^2)$ and $a_4 = a_4^{(0)} + O(\tau)$, it follows that $\eta \propto |\tau|^{1/2}$ in the ordered phase, leading to $\beta = 1/2$. To determine $\gamma$ and $\delta$, one again minimizes the Landau free energy, now including a nonzero external field, yielding $2a_2 \eta + 4 a_4 \eta^3 = h$. When this expression is evaluated at the critical point ($\tau = 0$), the $a_2$ coefficient vanishes, and one gets $h \propto \eta^3$; therefore $\delta = 3$ for the Landau model. Differentiating the same equation with respect to $h$ gives the susceptibility $\chi = \partial \eta/\partial h|_\tau = [2(a_2^{(0)} \tau + 6 a_4^{(0)} \eta^2)]^{-1}$. In the disordered phase ($\eta = 0$), this simplifies to $\chi \propto \tau^{-1}$. In the ordered phase, and for $h = 0$, $\eta^2 \propto \tau$, and we again find $\chi \propto \tau^{-1}$, leading to $\gamma = 1$ for the Landau model. These predictions for the critical exponents $\beta$, $\gamma$, and $\delta$ will be contrasted with numerical results in scalar-tensor theories in Sec.~\ref{sec:second}.

The rationale outlined above is well suited to the description of second-order phase transitions, in which case the order parameter $\eta$ varies continuously across phases. Still, the Landau model---grounded in analyticity arguments---can be adapted to describe first-order phase transitions, which are characterized by a discontinuous jump in $\eta$. One possibility involves introducing a cubic term into Eq.~\eqref{eq:Landau_ansatz}, thereby breaking the reflection symmetry of the free energy at $h=0$ and altering the relative depths of the local minima of $\mathcal{L}$. This asymmetry renders one minimum metastable, enabling a transition between phases. An alternative mechanism of realizing a first-order phase transition, while keeping the reflection symmetry of $\mathcal{L}$ at $h=0$, arises when the quartic coefficient $a_4$ becomes negative, and the free energy is stabilized by including higher-order terms in $\eta$. This can lead to multiple local minima with differing depths, with the transition between them being accompanied by a discontinuous shift in the order parameter. We will encounter such a scenario in Sec.~\ref{sec:first} below.

\section{Neutron stars in scalar-tensor theories}
\label{sec:NSinSTTs}

Scalarization can occur in a broad class of scalar-tensor theories (STTs) of gravity (see, e.g., Ref.~\cite{Andreou:2019ikc} for a discussion on the subclass of the Horndeski model that may be subject to the linear tachyonic instability that typically accompanies spontaneous scalarization). In this work, we will restrict our discussion to the class described by the following Einstein-frame action:
\begin{align} \label{eq:action_stts}
    S[g_{\mu\nu};\phi; \Psi_m] & = \frac{1}{16 \pi} \int \sqrt{-g} \left( R - 2 g^{\mu \nu} \partial_\mu \phi \partial_\nu \phi  \right) d^{4}x \nonumber \\
    & + S_{m}[\Psi_{m}, A(\phi)^{2} g_{\mu \nu}]  \, , 
\end{align}
where $R$ is the Ricci scalar, $\Psi_{m}$ collectively represents the matter fields, which are described by the action $S_{m}$, and universally couple to the conformally rescaled, Jordan-frame metric $\tilde{g}_{\mu\nu} = A(\phi)^2 g_{\mu\nu}$. Here, $A(\phi)$ is a free function of the scalar field that specifies the theory under consideration. For suitable choices of $A(\phi)$, this model allows for spontaneous scalarization of sufficiently compact neutron stars, although not of black holes \cite{Damour:1993hw,Sotiriou:2011dz}. 
Two one-parameter models will be adopted in what follows, namely
\begin{equation} \label{eq:models1and2}
    A_1(\phi)  =  e^{\frac{1}{2} \beta \phi^{2}} 
    \quad
    \textrm{and}
    \quad
    A_2(\phi) = [\cosh(\sqrt{3} \beta \phi)]^{1/3\beta}, 
\end{equation}
with $\beta$ being a constant. A discussion on their motivation can be found in Sec.~IIIB of  Ref.~\cite{Mendes:2016fby}.

We consider static, spherically symmetric neutron stars (NSs) in these classes of models. The spacetime is described by the line element
\begin{equation}\label{met_estrela}
    ds^2 = - e^{\nu(r)} dt^{2} + \left(1 - \frac{2 \mu(r)}{r} \right)^{-1} dr^{2} + r^{2} (d \theta^{2} + \sin{\theta}^{2} d \varphi^{2}) ,
\end{equation}  
and the star is modeled as a perfect fluid distribution with (Jordan-frame) energy-momentum tensor
\begin{equation}
    \tilde{T}^{\mu\nu} = (\tilde{\epsilon} + \tilde{p}) \tilde{u}^\mu \tilde{u}^\nu + \tilde{p} \tilde{g}^{\mu\nu},
\end{equation}
where $\tilde{u}^\mu$ denotes the four-velocity of fluid elements and $\tilde{\epsilon}$ and $\tilde{p}$ denote the energy density and pressure measured by comoving observers, which are assumed to be related by some predetermined, one-parameter equation of state (EoS) of the form $\tilde{p} = \tilde{p}(\tilde{\epsilon})$. For the numerical results presented in this work, we employ a piecewise polytropic approximation to the SLY9 EoS, with continuous speed of sound, following the parametrization provided in Ref.~\cite{OBoyle:2020qvf}. 

With the previous assumptions, the field equations derived from Eq.~\eqref{eq:action_stts} can be written as a set of ordinary differential equations for the state vector $[\mu(r), \nu(r), p(r), \phi(r), \psi(r)]$, with $\psi \equiv d\phi/dr$, which can be found, e.g., in Ref.~\cite{Damour:1993hw}.
These equations are numerically integrated by standard methods, subject to the boundary conditions $\mu(0) = 0$ (regularity), $\lim_{r\to\infty}\nu(r) = 0$ (asymptotic flatness), $p(0) = p_c$, $\psi(0) = 0$ (regularity), and $\lim_{r\to\infty}\phi(r) = \phi_0$ (an externally determined asymptotic scalar field, possibly of cosmological origin). The stellar radius $R_s$ is defined through $p(R_s) = 0$.

For the upcoming discussion, three global quantities will play an important role. The first is the stellar (ADM) mass $M$, which can be conveniently computed from the limit $M = \lim_{r\to \infty} \mu(r)$. The total baryon number is obtained from the (Jordan frame) number density $\tilde{n}=\tilde{n}(\tilde{p})$ as (see, e.g.,~\cite{Mendes:2016fby} for details),
\begin{equation}
    N = \int_0^{R_s} 4 \pi \tilde{n} A^3 r^2 (1 - 2\mu/r)^{-1/2} dr.
\end{equation}
Finally, the scalar charge $Q$ is defined as the scalar monopole moment of the source, such that
\begin{equation} \label{eq:Q}
    \phi(r) = \phi_0 + \frac{Q}{r} + O(r^{-2}), \qquad \textrm{for $r\gg R_s$}.
\end{equation}

Fixing the asymptotic value of the scalar field, $\phi_0$, one can generate a one-parameter sequence of equilibrium solutions by changing the value of the central pressure, $p_c$, or, equivalently, of the central number density, $\tilde{n}_c$, which consequently changes the total mass $M$, baryon number $N$, and scalar charge $Q$ of the star. If $\phi_0$ is allowed to change, then a two-parameter sequence of equilibrium solutions is produced.

\section{Scalarization as a phase transition}
\label{sec:SSPT}

\subsection{Basic elements}

To describe spontaneous scalarization as a phase transition, it is natural to consider the baryon number $N$, or, alternatively, the baryonic mass $M_b = m_n N$, where $m_n$ is the neutron rest mass, as determining the control parameter ($\tau$) of the system, since these quantities remain constant during dynamical processes governed by internal forces (i.e., with no mass injection). Specifically, if $M_{b,c}$ denotes the baryon mass at the critical point, then one can define $\tau = (M_{b,c} - M_b)/M_{b,c}$ (adjusting an overall sign if needed).

It is also natural to consider the scalar charge $Q$ as the order parameter ($\eta$), and the asymptotic scalar field, $\phi_0$, as the external field ($h$). In cases where spontaneous scalarization is well described by a continuous (``second-order'') phase transition, the general picture is, therefore, the following. In the absence of an external field ($\phi_0 =0$), and for $\tau > 0$, only one equilibrium solution exists, with $Q=0$. Past the critical point ($\tau < 0$), the $Q = 0$ solution becomes unstable under scalar field perturbations \cite{Harada:1997mr} and two new solutions appear, with $Q = \pm q \neq 0$. These are stable and constitute the true ground state of the system. We will see in what follows that this picture is often, but not always, a suitable one.

To access the validity of the Landau model, the next step is to find a candidate for the Landau free energy of the system. 
If one considers small changes among nearby equilibrium solutions, characterized by an infinitesimal variation $\delta \phi_0$ in the asymptotic scalar field and $\delta N$ in the total baryon number, it can be shown that the resulting change in the total mass (energy) of the system is given by:
\begin{equation}
    \delta M = - Q \delta \phi_{0} + \mu_\text{eff} \delta N \, ,
\end{equation}
where $\mu_\text{eff} = \partial M/\partial N | _{\phi_0}$ is an effective chemical potential \cite{Harada:1998ge}, and
\begin{equation} \label{eq_Q_dm_dphi0}
    Q = -\left.  \frac{\partial M}{\partial \phi_0}\right|_N
\end{equation}
can be shown to be equal to the scalar charge defined in Eq.~\eqref{eq:Q} (see Appendix A of Ref.~\cite{Damour:1992we}).

Therefore, the total energy $M$ has $(\phi_0,N)$ as natural variables. On the other hand, a suitable candidate for the Landau free energy should have the order parameter (the scalar charge $Q$) as a natural variable. 
From Eq.~\eqref{eq_Q_dm_dphi0}, one recognizes that $Q$ and $\phi_0$ are conjugate variables. This suggests the Legendre-transformed free energy \cite{Damour:1996ke}
\begin{equation} \label{eq_Leg_transf}
    m(Q) = M + Q \phi_0
\end{equation}
as a candidate for the Landau free energy, as it takes $Q$ and $N$ as its natural variables. Indeed, an infinitesimal variation $\delta Q$ in the scalar charge and $\delta N$ in the total baryon number induces a corresponding variation 
\begin{equation}
    \delta m = \phi_0 \delta Q + \mu_\text{eff} \delta N
\end{equation}
in this free energy.
Importantly, it has been verified by construction in the literature that $m$ typically satisfies the criteria for a Landau free energy, specially the fact that the state of the system corresponds to a global minimum of $m$ with respect to $Q$ \cite{Khalil:2019wyy,Khalil:2022sii}. Examples, as well as counterexamples, of this behavior will be discussed below.

\begin{figure}[tb]
     \includegraphics[width=0.95\linewidth]{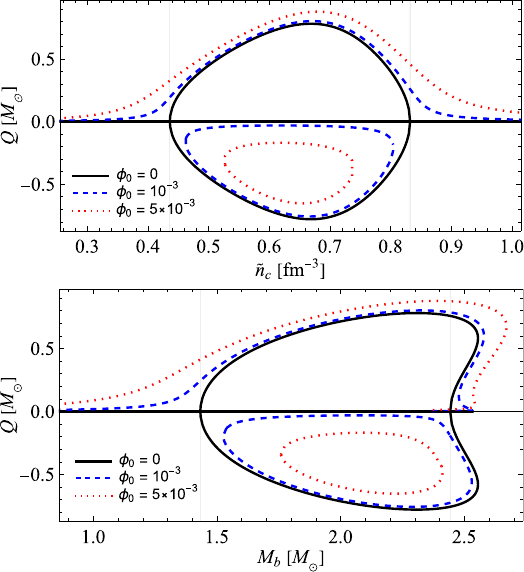}
     \caption{Scalar charge as a function of the central number density (upper panel) and of the total baryon mass (bottom panel) for model 1 with $\beta = -5$.
     Three values of the external field are considered, $\phi_0 \in \{0,10^{-3},5\times 10^{-3}\}$. Vertical gray lines indicate the two threshold points marking the beginning and end of the sequence of scalarized solutions.}
     \label{fig:QvsMb}
\end{figure}

\subsection{The ``standard'' scenario: Scalarization as a second-order phase transition}
\label{sec:second}

\begin{figure*}[t]
     \centering
     \includegraphics[width=\linewidth]{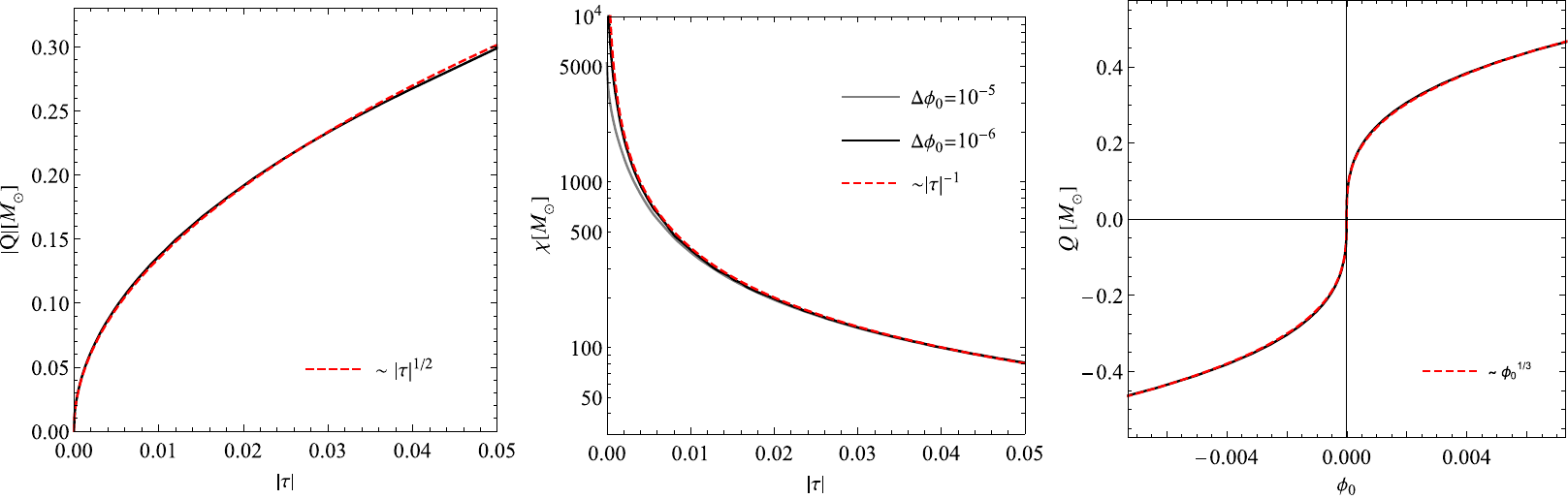}
     \caption{Critical exponents for spontaneous scalarization around the low-mass critical point in model 1 with $\beta = -5$. The left panel displays $Q$ as a function of the (absolute value of the) control parameter $\tau = (M_{b,1}^* - M_b)/M_{b,1}^*$. The middle panel shows the scalar susceptibility as a function of $|\tau|$, obtained according to Eq.~\eqref{eq:chiapprox}, with $\Delta \phi_0 \in \{10^{-5}, 10^{-6}\}$. The right panel shows the order parameter $Q$ as a function of the external field $\phi_0$ at the critical point $M_b = M_{b,1}^*$. Red dashed curves in all panels indicate best fits consistent with expectations from the Landau model.}
     \label{fig:criticalexponents}
\end{figure*}

Let us begin by revisiting the most extensively studied case in the literature, characterized by the coupling function $A_1(\phi)$ in Eq.~\eqref{eq:models1and2} (which we will refer to as model 1). To illustrate its key features, we set $\beta = -5$. 

Figure \ref{fig:QvsMb} shows the scalar charge (order parameter) in terms of both the central number density $\tilde{n}_c$ (upper panel) and the baryon mass $M_b$ (bottom panel). Two threshold points mark the beginning and end of the scalarization sequence, occurring at baryon masses $M_{b,1}^*\approx 1.433 M_\odot$ and $M_{b,2}^* \approx 2.443 M_\odot$. The non-monotonic relationship between $M_b$ and $\tilde{n}_c$ gives rise to the distinct features observed near the rightmost (higher-mass) critical point in the two plots, an issue that we explore below in Sec.~\ref{sec:first}. 

Critical exponents from the Landau model are contrasted with explicit computations near the low-mass critical point in Fig.~\ref{fig:criticalexponents}.
In the left panel, we zoom in on the low-mass critical point in Fig.~\ref{fig:QvsMb}, showing that $Q \propto |\tau|^\beta$, with $\beta \approx 1/2$, accurately describes the behavior of $Q(\tau,h=0)$ around $\tau = (M_{b,1}^* - M_b)/M_{b,1}^* = 0$. The middle panel presents the scalar susceptibility, defined as $\chi = \partial Q/\partial \phi_0|_{M_b}$, for $\phi_0 = 0$, as a function of the control parameter in the scalarized phase. To compute this quantity, in practice, we approximate it as
\begin{equation}\label{eq:chiapprox}
\chi(\phi_0 = 0,M_b) \approx \frac{Q(\Delta \phi_0, M_b)-Q(0, M_b)}{\Delta \phi_0},
\end{equation}
showing results for $\Delta \phi_0 \in \{10^{-5}, 10^{-6}\}$ in the plot. A reference curve proportional to $|\tau|^{-1}$ is also included, showing that $\gamma \approx 1$ in the scalarized phase. 
Finally, the right panel of Fig. \ref{fig:criticalexponents} displays $Q$ as a function of $\phi_0$, for $M_b = M_{b,1}^*$, showing that $\delta \approx 3$ provides a good fit for the relationship between the order parameter ($Q$) and the external field ($\phi_0$).

\begin{figure}[htb]
     \centering
     \includegraphics[width=0.95\linewidth]{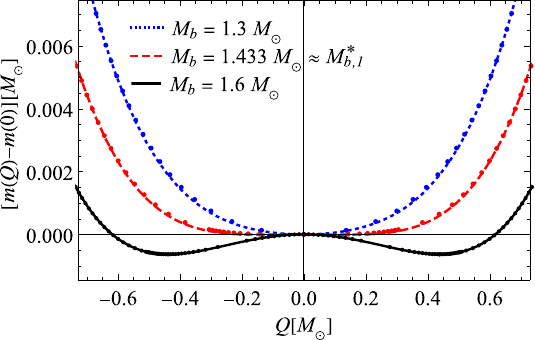}
     \caption{$m(Q,M_b)$ as a function of $Q$ for model 1 with $\beta = -5$ and for a few choices of the baryon mass $M_b$ of the star, close to the critical point at $M_{b,1}^* \approx 1.433 M_\odot$. Dots represent results of numerical calculations, whereas curves represent best fits of the quartic form \eqref{eq:expansion}. In this case, spontaneous scalarization is well described by the minimal Landau model.}
     \label{fig:effpot1}
\end{figure}

As expected, around the low-mass critical point the quantity $m(Q,M_b)$ defined in Eq.~\eqref{eq_Leg_transf} is well approximated by a quartic polynomial, of the form
\begin{equation} \label{eq:expansion}
    m(Q, M_b) = a_0(M_b) + a_2(M_b) Q^{2} + a_4(M_b) Q^4,
\end{equation}
with $a_2$ changing sign at the critical point. The behavior of $m$ as a function of $Q$ is illustrated in Fig.~\ref{fig:effpot1} for three values of the baryon mass. To construct this quantity, we evaluate Eq.~\eqref{eq_Leg_transf} for a fixed baryon mass while varying the asymptotic scalar field, following the procedure outlined in, e.g., Ref.~\cite{Khalil:2022sii}. For instance, for the baryon mass $M_b = 1.6 M_\odot > M_{b,1}^*$, three equilibrium solutions exist at $\phi_0 = 0$, corresponding to the three extrema in the respective curve in Fig.~\ref{fig:effpot1}. As $\phi_0$ increases, the scalar charge $Q$ grows for the two solutions with large $|Q|$ and decreases for the low-$|Q|$ solution (cf.~Fig.~\ref{fig:QvsMb}); the opposite behavior occurs when $\phi_0$ becomes increasingly negative. The entire curve is then traced by sweeping $\phi_0$ over a relevant range.
The fact that the free energy at zero external field is obtained in practice by varying $\phi_0$ is somewhat intriguing. Nevertheless, in the absence of a direct procedure to derive the effective free energy through a renormalization group analysis or a mean-field approximation from a fundamental action, this approach still provides a reasonable phenomenological construct.
Overall, we see that, at least for the case at hand, spontaneous scalarization conforms to the basic views of the Landau model. 

Before turning to different scenarios, it is worth commenting on the relevance of these results (for isolated neutron stars) to the phenomenon of dynamical scalarization occurring in multibody systems. Consider a binary system of two neutron stars with a wide separation, such that the interbody distance $r$ is much larger than the stellar radius, i.e., $\varepsilon = r/R_s \ll 1$. As explained in \cite{Damour:1992we}, one can define worldtubes around each neutron star of an intermediate length scale $r_{match} = \sqrt{rR_s}$ and solve the problem separately inside and outside these worldtubes, ensuring consistency at $r_{match}$. A significant simplification arises in the limit $\varepsilon \to 0$, since, when expressed in units of $R_s$, the matching radius scales as $r_{match}/R_s = \varepsilon^{-1/2} \to \infty$, while in units of $r$, it scales as $r_{match}/r = r \varepsilon^{1/2} \to 0$. In this limit, from the perspective of the outer problem, the worldtubes collapse to worldlines, where appropriate boundary conditions must be imposed. Meanwhile, from the perspective of the inner problem, the matching radius extends to infinity, effectively reducing the problem to that of isolated neutron stars with asymptotically prescribed boundary conditions.

In this framework, the asymptotic scalar field $\phi_0$, which originates cosmologically for an isolated neutron star, can instead be reinterpreted as a boundary condition imposed by the presence of a companion in a binary system. This perspective naturally leads to an explanation of dynamical scalarization, as encapsulated, for instance, in the feedback model proposed in Ref.~\cite{Palenzuela:2013hsa}. As two neutron stars, $A$ and $B$, approach one another, the asymptotic boundary condition for $A$ ($\phi_0^{(A)}$) is determined by the far-zone scalar field generated by $B$ --- which, to leading order, is given by Eq.~\eqref{eq:Q} --- and vice versa. Thus, the effective boundary condition $\phi_0^{(A,B)}$ for stars $A$ and $B$ can be approximately extracted from the fixed point of the system
\begin{equation}\label{eq:feedback}
    \phi_{0}^{(A)} = \phi_{0} + \frac{Q_{B}(\phi_{0}^{(B)})}{r}  \, , \qquad
    \phi_{0}^{(B)} = \phi_{0} + \frac{Q_{A}(\phi_{0}^{(A)})}{r}  \, .
\end{equation}

The upper panel of Fig.~\ref{fig:DS} shows the scalar charge as a function of the orbital separation distance for an equal-mass binary system with $M_{b,A} = M_{b,B} = 1.3 M_\odot$, computed using the feedback model described above and assuming an adiabatic evolution (see Ref.~\cite{Khalil:2022sii} for an alternative model that captures nonadiabatic features). While these stars would not scalarize in isolation, they acquire a significant scalar charge in a close binary system. In Sec.\ref{sec:other}, we contrast this scenario with a more unconventional case in which dynamical descalarization occurs instead.

\begin{figure}[thb]
     \centering
     \includegraphics[width=0.97\linewidth]{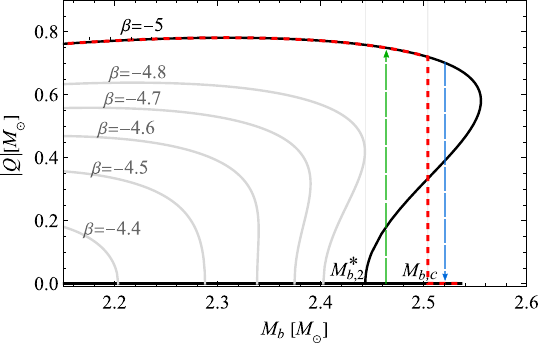}
     \caption{Absolute value of the scalar charge as a function of baryon mass for $\beta \in \{-4.4,-4.5,-4.6,-4.7,-4.8,-5 \}$ in model 1, around the high-mass critical point, at zero external field, $\phi_0=0$. As $\beta$ decreases, the phase transition shifts from second-order to first-order. Vertical gray lines indicate the threshold baryon mass for scalarization, $M_{b,2}^* \approx 2.443 M_\odot$, and the critical baryon mass $M_{b,c} \approx 2.504 M_\odot$, both for the $\beta = -5$ case. The red dashed line tracks the minimum energy solution, which corresponds to the ground state of the system. Green and blue vertical arrows indicate the approximate baryon masses for the dynamical simulations reported in Sec.~\ref{sec:nonlinear_dynamics}. }
     \label{fig:QvsMbvariousbeta}
\end{figure}

\section{Scalarization as a first-order phase transition}
\label{sec:first}

\subsection{Free energy analysis}

Let us now focus on the rightmost, higher-mass critical point seen in Fig.~\ref{fig:QvsMb}. Independently of the value of $\beta$, the $Q -\tilde{n}_c$ relation retains the qualitative features shown in the upper panel of Fig.~\ref{fig:QvsMb}. Specifically, for a given central density within the scalarization range, there are always three solutions at zero external field: one with $Q = 0$ and two with nonzero scalar charges of opposite sign. 

However, the behavior of the $Q -M_b$ relation around the high-mass critical point is much more sensitive to the choice of coupling parameter $\beta$, as illustrated in Fig.~\ref{fig:QvsMbvariousbeta}. For larger values of $\beta$, such as $\beta = -4.5$, which lie closer to the boundary of the scalarization region, the baryon mass $M_b$ varies monotonically with $\tilde{n}_c$ along the existence line of scalarized solutions, and consequently three solutions exist for a fixed baryon mass within the scalarization range. In such cases, similar features to those presented in Sec.~\ref{sec:second} would characterize the (second-order) phase transition around the higher-mass critical point. 

\begin{figure}[htb]
     \centering
     \includegraphics[width=0.95\linewidth]{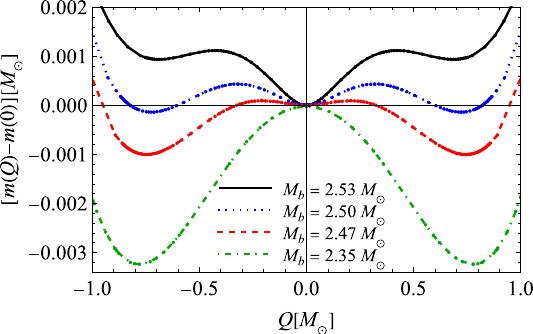}
     \caption{$m(Q,M_b)$ as a function of $Q$ for model 1 with $\beta = -5$ and for a few choices of the baryon mass $M_b$ of the star, close to the threshold value $M_{b,2}^* \approx 2.443 M_\odot$. Dots represent results of numerical calculations, while curves are linear interpolations between those points.}
     \label{fig:effpot2}
\end{figure}

However, as the $\beta = -5$ case illustrates, the baryon mass is typically not a monotonically increasing function of the central density along the sequence of scalarized solutions. As a result, for a fixed $M_b$ near the higher-mass critical point, more than three solutions may exist, changing the character of the phase transition. An interesting situation occurs when, as in the $\beta = -5$ case, the scalarization sequence itself has a turning point, but terminates before the turning point of the nonscalarized one. Then, for a fixed baryon mass larger, but close to that of the higher-mass threshold for scalarization (which we label as $M_{b,2}^*$), five equilibrium solutions exist\footnote{For the purposes of this discussion, we neglect a sixth unstable solution that would exist beyond the turning point of the nonscalarized sequence.} at zero external field: a (linearly) stable solution with $Q = 0$, two unstable solutions with $Q = \pm q_1$, and two (linearly) stable solutions with $Q = \pm q_2$ (with $|q_2|>|q_1|$). 
These solutions correspond to the extrema of the plots shown in Fig.~\ref{fig:effpot2}, which depict the free energy $m(Q,M_b)$ for a few values of the total baryon mass near $M_{b,2}^* \approx 2.443 M_\odot$. At these extremal points, the free energy $m$ reduces to the total mass $M$ (since $\phi_0 = 0$), allowing for a straightforward energy analysis. 

From Fig.~\ref{fig:effpot2} we see that, for a high baryon mass (e.g.~$M_b = 2.53M_\odot$), the $Q = 0$ solution is the true ground state of the system, while the $Q = \pm q_2$ solutions are metastable. However, as $M_b$ decreases, a critical point occurs at $M_{b,c} \approx 2.504 M_\odot$ where the $Q=0$ solution and the $Q = \pm q_2$ solutions have the same total mass. For $M_b < M_{b,c}$, the scalarized solutions become energetically favored over the $Q=0$ solution and thus constitute the true ground state of the system, while the $Q=0$ solution becomes metastable. At the threshold mass $M_{b,2}^*$, the concavity of $m(Q)$ at $Q=0$ changes, consistent with the fact that this solution undergoes a stability transition at that point. 

The simplest model that respects the symmetries of the problem while capturing the qualitative features described above extends the Landau ansatz \eqref{eq:Landau_ansatz} to 
\begin{equation} \label{eq:Landau_ansatz_6}
    \mathcal{L} = a_{2} \eta^{2}  + a_{4} \eta^{4} + a_6 \eta^6,
\end{equation}
with $a_6 > 0$. As discussed in Sec.~\ref{sec:pt}, a first-order phase transition would be characterized by a change in sign of the coefficient $a_4$. While this qualitatively aligns with the schematic behavior shown in Fig.~\ref{fig:effpot2}, we find that a sixth-degree polynomial does not provide an accurate fit for the full form of the free energy $m$, in contrast with the cases illustrated in Fig.~\ref{fig:effpot1}, which are well fitted by a quartic polynomial of the form \eqref{eq:expansion}.
However, this is not surprising since, in contrast to second-order phase transitions, which strongly rely on the system's behavior in the immediate vicinity of the critical point, first-order phase transitions involve larger excursions of the order parameter. Consequently, a low-order polynomial expansion may be insufficient, necessitating the inclusion of higher-order terms (of $\sim 18$th order, in our case) for an accurate representation. 

Before turning to analyze the dynamics of the first-order phase transition, we note that similar features would arise around the leftmost, low-mass critical point for sufficiently negative values of $\beta$, as recently observed in Ref.~\cite{Unluturk:2025zie}\footnote{We became aware of Ref.~\cite{Unluturk:2025zie} while the present work was being completed.}. For the EoS adopted in our study, when $\beta \lesssim -10.25$, then the baryon mass becomes a locally decreasing function of $\tilde{n}_c \gtrsim \tilde{n}_{c,1}$ (where $\tilde{n}_{c,1}$ denotes the low-mass threshold central density for scalarization) along the sequence of scalarized solutions. Then arguments similar to those discussed earlier imply that the phase transition at this point would be first-order rather than second-order. It is worth mentioning that model 1 with such low values of $\beta$ is ruled out by pulsar timing observations \cite{Damour:1996ke,Freire:2012mg,Anderson:2019eay}, but a similar feature can arise in massive scalar-tensor theories (as considered in Ref.~\cite{Unluturk:2025zie}), for which binary pulsar observations are significantly less constraining \cite{Ramazanoglu:2016kul,Degollado:2024oyo}.

\subsection{Nonlinear dynamics} \label{sec:nonlinear_dynamics}

In this section we report results from fully nonlinear simulations that show the dynamical processes of scalarization and descalarization in the first-order regime. For definiteness, we focus on the case of model 1 with $\beta=-5$ and $\phi_0=0$.

As initial data for scalarization, we take an equilibrium, nonscalarized configuration with baryon mass in the range $M^*_{b,2} < M_b < M_{b,c}$. Specifically, we assume $M_b = 2.462 M_\odot$, corresponding to a central number density $\tilde{n}_c = 0.854\, \textrm{fm}^{-3}$, as indicated by the green vertical arrow in Fig.~\ref{fig:QvsMbvariousbeta}.
In this case, we know in advance that the initial data is metastable: although it is stable under linear perturbations, it is not the true ground state of the system, as energetically favored scalarized solutions exist with the same baryon mass. 
To trigger the transition to the scalarized configuration, the system must be perturbed strongly enough to supersede the existing potential barrier between these states.
In practice, we initialize the scalar field with a nontrivial velocity profile $\dot{\phi}(t=0,r)$, which provides the system with the kinetic energy needed to overcome the barrier---see the Appendix for technical details. After perturbing the original configuration, we impose the field equation constraints to construct consistent initial data for time evolution. This perturbation slightly alters the original baryon mass to $M_b = 2.464 M_\odot$.

\begin{figure}[htb]
     \centering
     \includegraphics[width=1.0\linewidth]{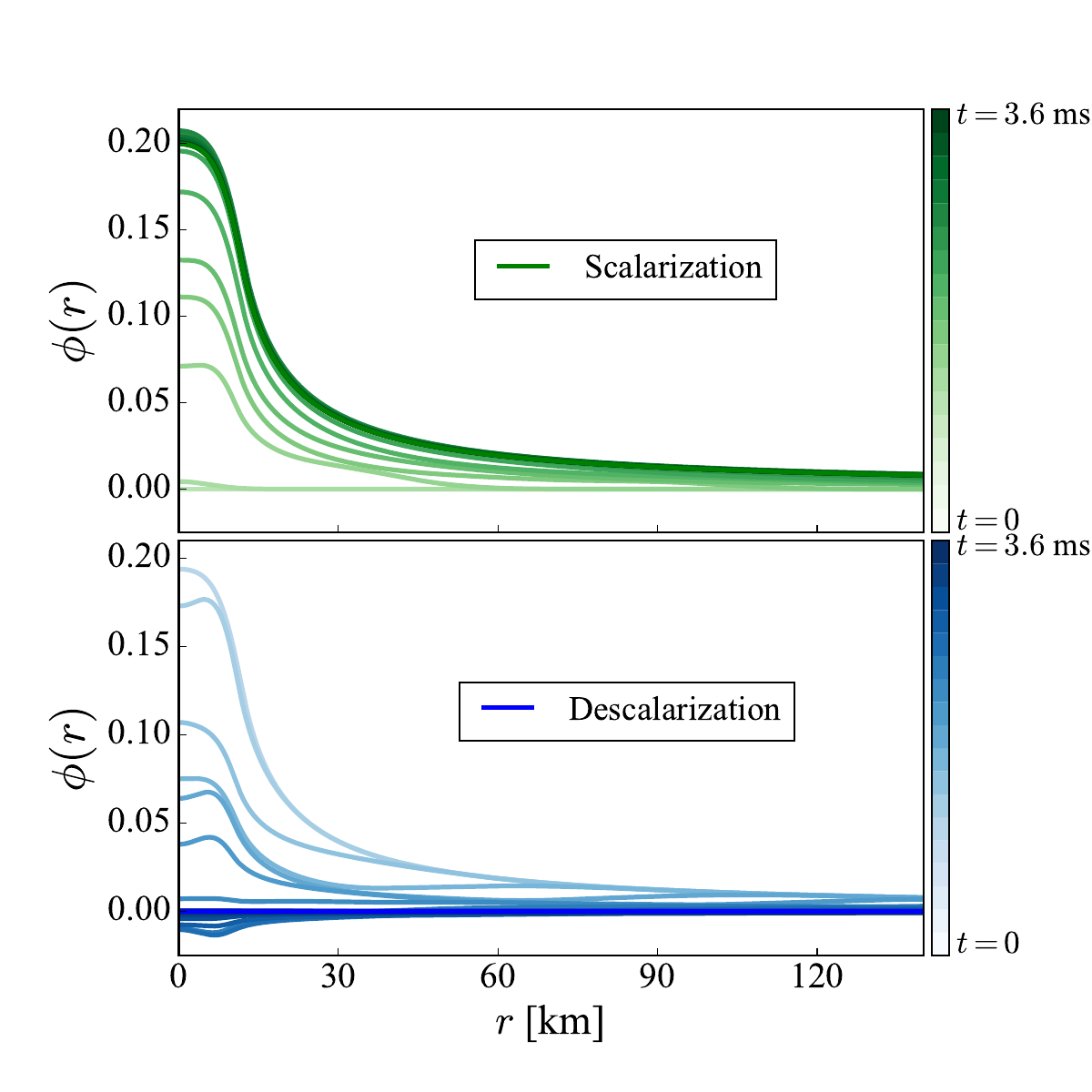}
     \caption{Time frames of the scalar field evolution during scalarization (upper panel) and descalarization (lower panel) processes in the first-order phase transition regime. Time ranges from $0$ to $\sim 3.6$ ms, with darker colors corresponding to later times.}
     \label{fig:snapshots}
\end{figure}

The upper panel of Fig. \ref{fig:snapshots} displays snapshots of the evolving radial scalar field profile over time, with lighter colors indicating earlier times. 
In the late stages, the system settles into a configuration consistent with a scalarized solution characterized by $\phi_c \approx 0.2005$, $\tilde{n}_c \approx 0.7103 \textrm{fm}^{-3}$,
and $M_b \approx 2.464 M_\odot$. The excess energy is quickly released in the form of scalar waves, through the excitation of a strongly damped $\phi$-mode \cite{Mendes:2018qwo}.
Ideally, this scalarization process should keep the baryon mass constant. In practice, however, numerical errors induce a slight increase in $M_b$, with the growth rate decreasing as numerical resolution increases---quantitative details are reported in the Appendix. Figure \ref{fig:time_evolution} displays, in green, the time evolution of the central value of the scalar field, $\phi_c$ (upper panel), as well as the central value of the number density, $\tilde{n}_c$ (lower panel).

\begin{figure}[t]
     \centering
     \includegraphics[width=1.05\linewidth]{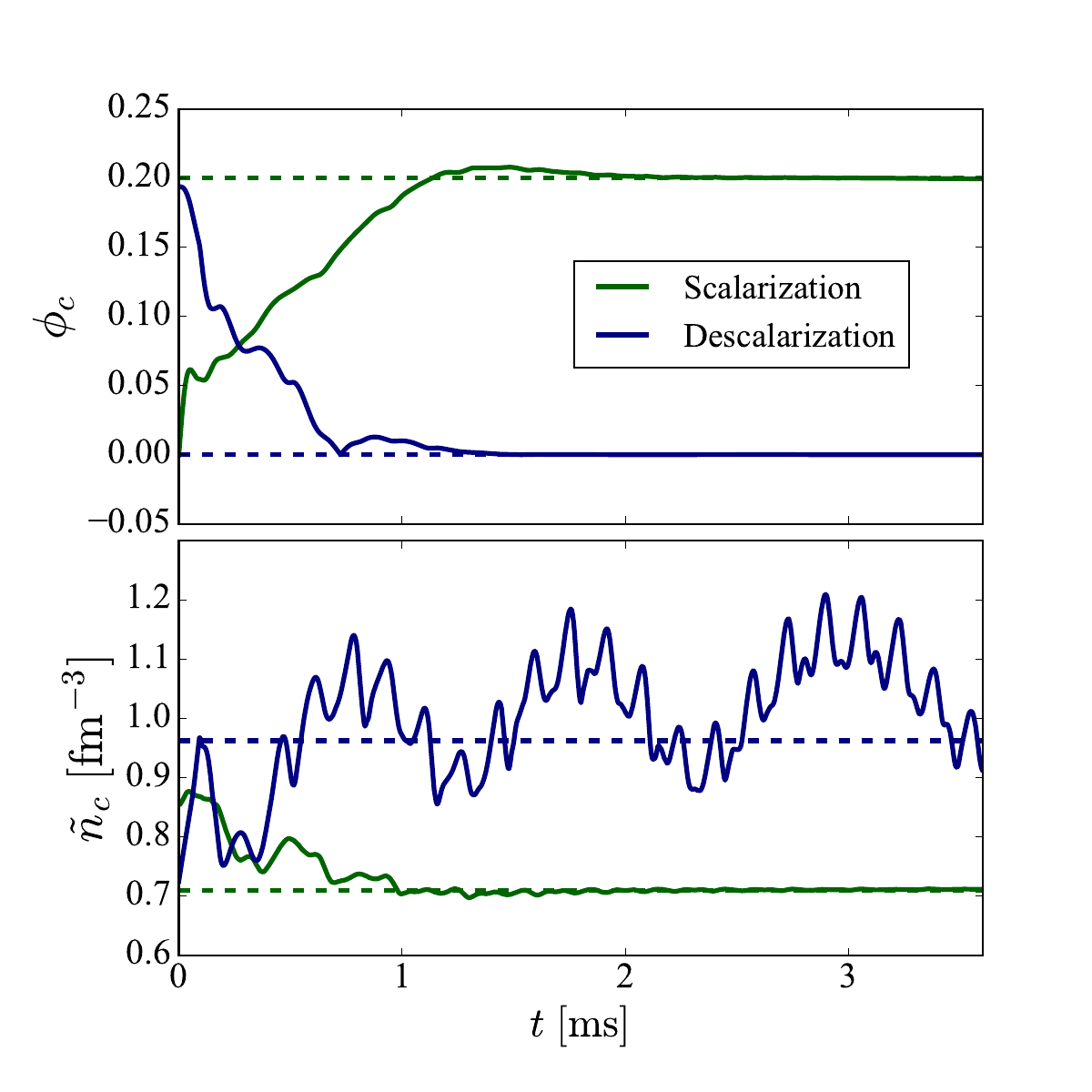}
     \caption{Time evolution of central values of the scalar field, $\phi_c$ (upper panel), and number density, $\tilde{n}_c$ (lower panel), during scalarization (green) and descalarization (blue) processes in the first-order regime. Dotted lines indicate $\phi_c$ and $\tilde{n}_c$ for static equalibrium configurations with the same baryon mass of the initial perturbed solutions, but energetically favored over them. In both cases, the initial perturbation is able to move the system away from the initial metastable state towards its true ground state.}
     \label{fig:time_evolution}
\end{figure}

To illustrate the descalarization process, we begin with an equilibrium scalarized configuration with baryon mass in the range $M_{b,c} < M_b < M_{b,\text{max}}$, where $M_{b,\text{max}}$ denotes the maximum baryon mass of nonscalarized solutions. Specifically, we take $M_b = 2.506 M_\odot$, as indicated by the blue vertical arrow in Fig.~\ref{fig:QvsMbvariousbeta}, which corresponds to $\phi_c = 0.1940$ and $\tilde{n}_c = 0.7254 \, \textrm{fm}^{-3}$. 
Again, in this case, the initial data is metastable: it is linearly stable but does not represent the system's true ground state. To induce the transition, a sufficiently strong perturbation is required for the system to overcome the potential barrier and reach the energetically favored configuration.
Again, we induce the transition by supplying the necessary kinetic energy to overcome the barrier. In practice, this is achieved by initializing a nontrivial velocity profile for the fluid, which raises the baryon mass of the initial data to $M_b = 2.521 M_\odot$.

The lower panel of Fig.~\ref{fig:snapshots} shows the expected decay of the initial scalar field profile to zero, with darker colors corresponding to later times. Correspondingly, Fig.~\ref{fig:time_evolution} shows, in blue, the time evolution of the central value of the scalar field, $\phi_c$ (upper panel), and the central value of the number density, $\tilde{n}_c$ (lower panel). The system oscillates around a configuration consistent with a nonscalarized solution with 
$\tilde{n}_c = 0.9625 \, \textrm{fm}^{-3}$ and $M_b = 2.521 M_\odot$. The excess energy that is not radiated in the form of scalar waves feeds strong (undamped) oscillations of the density profile, characterized by a superposition of a low-frequency $F$ mode ($f \sim 1.1$ kHz) and higher-frequency components, predominantly the $H_1$ mode ($f \sim 6.2$ kHz).

We remark that simulations similar to those presented here were carried out in Ref.~\cite{Kuan:2022oxs}, which discussed the process of accretion-induced descalarization in massive STTs. A key difference in their study is that, in the scenarios they considered, the nonscalarized branch possessed a higher total mass than the scalarized one. Thus, by starting with a scalarized configuration and gradually increasing its mass through an accretion-like process, the system eventually transitioned to the nonscalarized branch. 
In contrast, our simulations --- enabled by a careful treatment of the constraint equations --- demonstrate that, with a sufficiently strong perturbation, the transition can occur well before the maximum-mass solution is reached.

\section{A more unconventional scenario: Negative scalar susceptibility} 
\label{sec:other}

Since the discovery of spontaneous scalarization by Damour and Esposito-Farèse \cite{Damour:1993hw}, many instances have been found in the literature of theories exhibiting the same basic phenomenology: For zero asymptotic scalar field ($\phi_0 = 0$), the field equations admit a trivial solution with $\phi = 0$, which becomes unstable under linear scalar field perturbations beyond a certain threshold. At this point, new branches of solutions with a nontrivial scalar field profile emerge, which are linearly stable and energetically favored over the now-unstable trivial solution. However, the presence of these essential features does not necessarily imply that the nature of the phase transition is the same as discussed in Sec.~\ref{sec:second} or \ref{sec:first}. Neither does it imply that the Landau model can be straightforwardly adjusted to describe the system under consideration.
In this section, we aim to illustrate this point. For this purpose, we consider the case of the coupling function $A_2(\phi)$ defined in Eq.~\eqref{eq:models1and2} (which we will refer to as model 2), with $\beta = 100$. 

\begin{figure}[htb]
     \centering
     \includegraphics[width=0.95\linewidth]{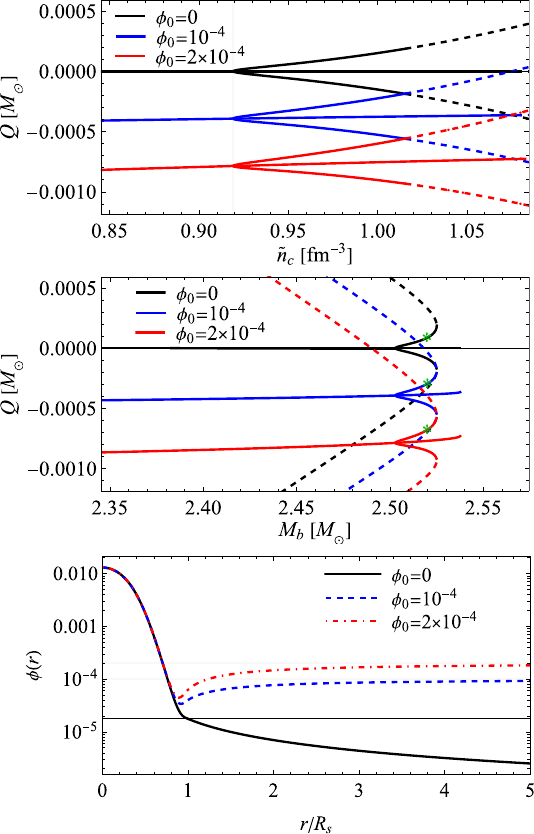}
     \caption{Scalar charge as a function of the central number density (upper panel) and of the total baryon mass (middle panel) for model 2 with $\beta = 100$.
     Three values of the external field are considered, $\phi_0 \in \{0,10^{-4},2\times 10^{-4}\}$. The bottom panel shows the scalar field profile for three solutions with $M_b = 2.52 M_\odot$, but varying asymptotic scalar field $\phi_0$ (marked with with green stars in the middle panel).}
     \label{fig:Qb100}
\end{figure}

The coupling functions $A_1$ and $A_2$ exhibit the same behavior for small scalar fields, namely, $A(\phi) = 1 + \beta \phi^2/2 + O(\phi^4)$, differing only at higher orders. As a consequence, the trivial $\phi = 0$ solution is linearly unstable in both theories for the same range of stellar properties \cite{Mendes:2014ufa}. When $\beta < 0$, these models differ only in the quantitative properties of scalarized solutions. However, for $\beta >0$, the distinction is more pronounced. In model 1, scalarized solutions for $\beta >0$ are typically unstable and a premature gravitational collapse follows the tachyonic instability of the $\phi = 0$ solution
\cite{Mendes:2016fby,Palenzuela:2015ima}. In contrast, in model 2, scalarized solutions can be stable and energetically favored over the $\phi = 0$ solution \cite{Mendes:2016fby}.

\begin{figure}[htb]
     \centering
     \includegraphics[width=0.95\linewidth]{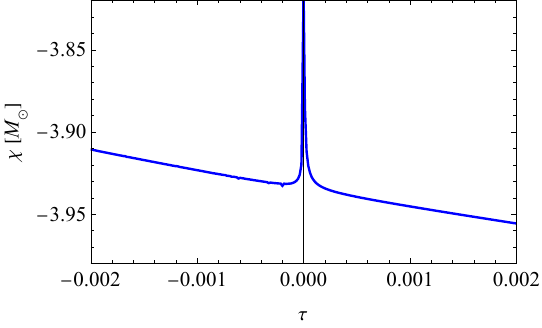}
     \caption{Scalar susceptibility at zero external field as a function of $\tau = (M_{b,c} - M_b)/M_{b,c}$, where $M_{b,c} \approx 2.5031 M_\odot$ is the critical mass for scalarization in model 2 with $\beta = 100$. $\chi$ was computed according to Eq.~\eqref{eq:chiapprox}, using $\Delta \phi_0 = 10^{-5}$.}
     \label{fig:chib100}
\end{figure}

Figure \ref{fig:Qb100} shows the scalar charge as a function of both the central density and the baryon mass for model 2 with $\beta = 100$. For $\phi_0 = 0$, there is a critical point that marks both the onset of instability of the $Q = 0$ solution and the emergence of new branches of scalarized solutions---consistent with the standard lore. A key difference from the $\beta < 0$ case lies in the effect of a nonzero asymptotic boundary condition: As $\phi_0$ increases, the scalar charge tends to decrease rather than increase, and the scalar susceptibility $\chi = \partial Q/\partial\phi_0|_{M_b}$ turns out to be a mostly negative quantity; see Fig.~\ref{fig:chib100}. This behavior can be understood from the bottom panel of Fig. \ref{fig:Qb100}, which depicts the radial scalar field profile for solutions with the same baryon mass but different values of $\phi_0$: Although the central scalar field increases with $\phi_0$, the rate at which the solution approaches $\phi_0$ as $r\to\infty$---which determines the scalar charge via Eq.~\eqref{eq:Q}--- decreases, eventually becoming negative (as $\phi_0$ is approached from below). 

A first consequence of this behavior is that the free energy $m(Q) = M + Q\phi_0$ no longer satisfies the criteria for a Landau free energy. Indeed, for a fixed baryon mass below the scalarization threshold, $m(Q)$ exhibits a local maximum at $Q=0$, instead of a local minimum, which can be traced to the fact that the $Q\phi_0$ term contributes negatively to the free energy $m(Q)$. Note, however, that the fact that the concavity of $m(Q)$ is not necessarily linked to the stability of the solution at an extremal point also manifests in the $\beta < 0$ case: a $Q=0$ solution past the turning point in the $M-\tilde{n}_c$ diagram is unstable under radial perturbations, yet it can be verified by explicit construction that it is a local minimum of $m(Q)$.

\begin{figure}[htb]
     \centering
     \includegraphics[width=0.95\linewidth]{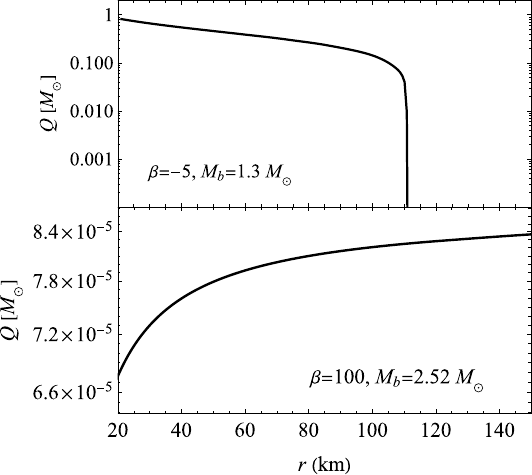}
     \caption{Dynamical scalarization in model 1 with $\beta = -5$ (upper panel) and descalarization in model 2 with $\beta = 100$ (bottom panel), as predicted by the feedback model \eqref{eq:feedback}. The scalar charge is plotted as a function of the orbital separation distance for an equal-mass binary system. In the upper panel, the individual baryon masses ($M_b = 1.3 M_\odot$) are below the threshold for spontaneous scalarization, meaning the stars are unscalarized in isolation but acquire a scalar charge beyond a critical orbital separation. In the bottom panel, the stars ($M_b = 2.52 M_\odot$ each) are initially scalarized, but undergo descalarization as the orbital separation decreases.}
     \label{fig:DS}
\end{figure}

Perhaps even more interesting are the implications of a negative scalar susceptibility for dynamical scalarization. As discussed in Sec.~\ref{sec:second}, valuable insights into this effect arise from considering that the presence of a companion effectively alters the ``asymptotic'' scalar field experienced by a star, with the leading-order correction being proportional to the scalar charge, as encapsulated in the feedback model (\ref{eq:feedback}). If $Q$ increases with $\phi_0$, a positive feedback mechanism enhances the scalar charge as the stars move closer together. Conversely, if $Q$ decreases with increasing $\phi_0$, a negative feedback effect should suppress the scalar charge as the binary system coalesces---as anticipated in Ref.~\cite{Mendes:2019zpw}. This behavior is illustrated in the lower panel of Fig.~\ref{fig:DS} for model 2 with $\beta = 100$, which shows the scalar charge as a function of orbital separation, as predicted by the feedback model. The stars, each with a baryon mass of $2.52 M_\odot$, carry a nontrivial scalar charge in isolation, which gradually decreases with decreasing orbital separation.

\section{Conclusions} \label{sec:conclusions}

Spontaneous scalarization has traditionally been described as a second-order phase transition. However, as similar phenomenology emerges across a broader range of models, it is worth reassessing the foundations and potential limitations of this interpretation. At present, the phase-transition nature of spontaneous scalarization is understood not from a fundamental perspective but primarily through the lens of Landau's phenomenological theory. In this paper, we critically examine this framework --- its successes, as well as its shortcomings --- through a few selected examples.

As shown in Fig.~\ref{fig:criticalexponents}, the critical exponents for spontaneous scalarization often align closely  with the predictions of the Landau model; moreover, the Legendre-transformed free energy $m(Q,M_b) = M + Q\phi_0$ often meets the basic criteria for a Landau free energy, as illustrated in Fig.~\ref{fig:effpot1}.

A slight departure from the traditional view was analyzed in Sec.~\ref{sec:first}, where the phase transition near the high-mass critical point in model 1 with $\beta = -5$ was shown to be of first rather than second order (cf.~Figs.~\ref{fig:QvsMbvariousbeta} and \ref{fig:effpot2}). In this case, the stability properties of the relevant solutions remain well captured by the form of the free energy $m(Q)$. Additionally, although the minimal extension in Eq.~\eqref{eq:Landau_ansatz_6} is insufficient to fully describe the profile of $m(Q)$, the fundamental reasoning behind interpreting a first-order phase transition within Landau’s framework still holds. 

Moreover, we have explored, through full nonlinear numerical simulations, the dynamical transition between metastable states (either scalarized or nonscalarized) and the true ground state of the system. Since the initial configurations are linearly stable, a sufficiently strong perturbation is required to trigger the transition. Figures \ref{fig:snapshots} and \ref{fig:time_evolution} show representative simulations exhibiting both scalarization and descalarization processes in the first-order regime. 
This is akin to the ``gravitational transition'' found in studies of accretion-induced descalarization in massive STTs \cite{Kuan:2022oxs}. 
As a topic for future research, it would be interesting to confirm, by the explicit construction of a suitable Landau free energy, that a first-order phase transition also characterizes the gravitational transition of black holes in some scalar-Gauss-Bonnet theories, such as the models explored in Ref.~\cite{Doneva:2021tvn,Doneva:2023kkz,Pombo:2023lxg}. 

Finally, we examined a more unconventional scenario (model 2 with $\beta = 100$) which, while exhibiting the basic features of what is typically called ``spontaneous scalarization'', has the distinctive characteristic that the scalar charge (order parameter) decreases with increasing $\phi_0$ (external field), rather than increasing. This results in a negative scalar susceptibility, with interesting implications for the process of dynamical scalarization. Indeed, as a binary system coalesces in this theory, the scalar charge is suppressed instead of enhanced by the approaching companion (cf.~Fig.~\ref{fig:DS}).

The unconventional behavior of the scalar charge with respect to the external scalar field could be considered a curiosity of this model; however, it ultimately undermines the favorable properties of our candidate $m$ for a Landau free energy. Recall that the space of solutions in scalar-tensor theories is two-dimensional, and can be described by pairs of variables such as $(p_c, \phi_0)$ or $(M_b, Q)$. The construction of $m(Q)$ for a fixed baryon mass $M_b$ relies on sweeping through a relevant range of $\phi_0$. Thus, curiously, in the traditional approach, the profile of the Landau free energy at zero external field is effectively constructed by varying $\phi_0$ itself. The negative scalar susceptibility in this model, where $Q$ decreases with increasing $\phi_0$, then ends up impairing the desirable features of $m(Q)$. In particular, the stability properties of equilibrium solutions at $\phi_0 = 0$ are no longer captured by the concavity of $m(Q)$, and the prescription for constructing a reasonable Landau free energy in this context remains an open issue.

\acknowledgments
The authors thank Daniel Stariolo for insightful discussions on the topic of phase transitions. J.M. acknowledges support from the Brazilian funding agency CAPES. R.M. acknowledges support from CNPq and from the FAPERJ grant SEI-260003/013339/2024. N.O. acknowledges support from the UNAM-PAPIIT grant IA101123. Numerical simulations were performed at the \textsc{LAMOD} facility at ICN, UNAM.


\appendix
\section{Numerical setup}

The results from the nonlinear simulations presented in this work were obtained using the code previously described in Refs. \cite{Mendes:2016fby,Mendes:2021fon}. Those references provide detailed accounts of the initial data construction, time evolution scheme, gauge choices, boundary conditions, vacuum treatment, shock capturing, and convergence tests. Here, we focus on the perturbation techniques specific to this work as well as on the simulation performance depending on numerical resolution.

For the process of scalarization discussed in Sec.~\ref{sec:nonlinear_dynamics}, the initial nonscalarized state must be perturbed strongly enough to overcome the potential barrier and transition to the scalarized ground state. This is implemented by setting up a non-zero initial velocity profile for the scalar field, which induces the formation of a nontrivial scalar field configuration. Specifically, we use the Gaussian function $\dot{\phi}(t=0,r) = \epsilon  e^{-2r^2/R_s^2} $, where $R_s$ is the stellar radius, and the amplitude $\epsilon$ characterizes the perturbation strength. For production runs we take $\epsilon = -0.5$.

In the case of descalarization, we instead perturb the fluid's radial velocity, which is initially zero in the equilibrium configuration. We take $v(t=0,r) = \epsilon r/R_s$, where $\epsilon < 0$ characterizes the perturbation strength, which obeys the necessary regularity conditions at the origin. For the simulations presented in Figs.~\ref{fig:snapshots} and \ref{fig:time_evolution}, we use $\epsilon = -0.1$. This type of perturbation provides the necessary kinetic energy for the system to overcome the potential barrier and transition to its descalarized ground state.

\begin{figure}[b]
     \centering
     \includegraphics[width=1.05\linewidth]{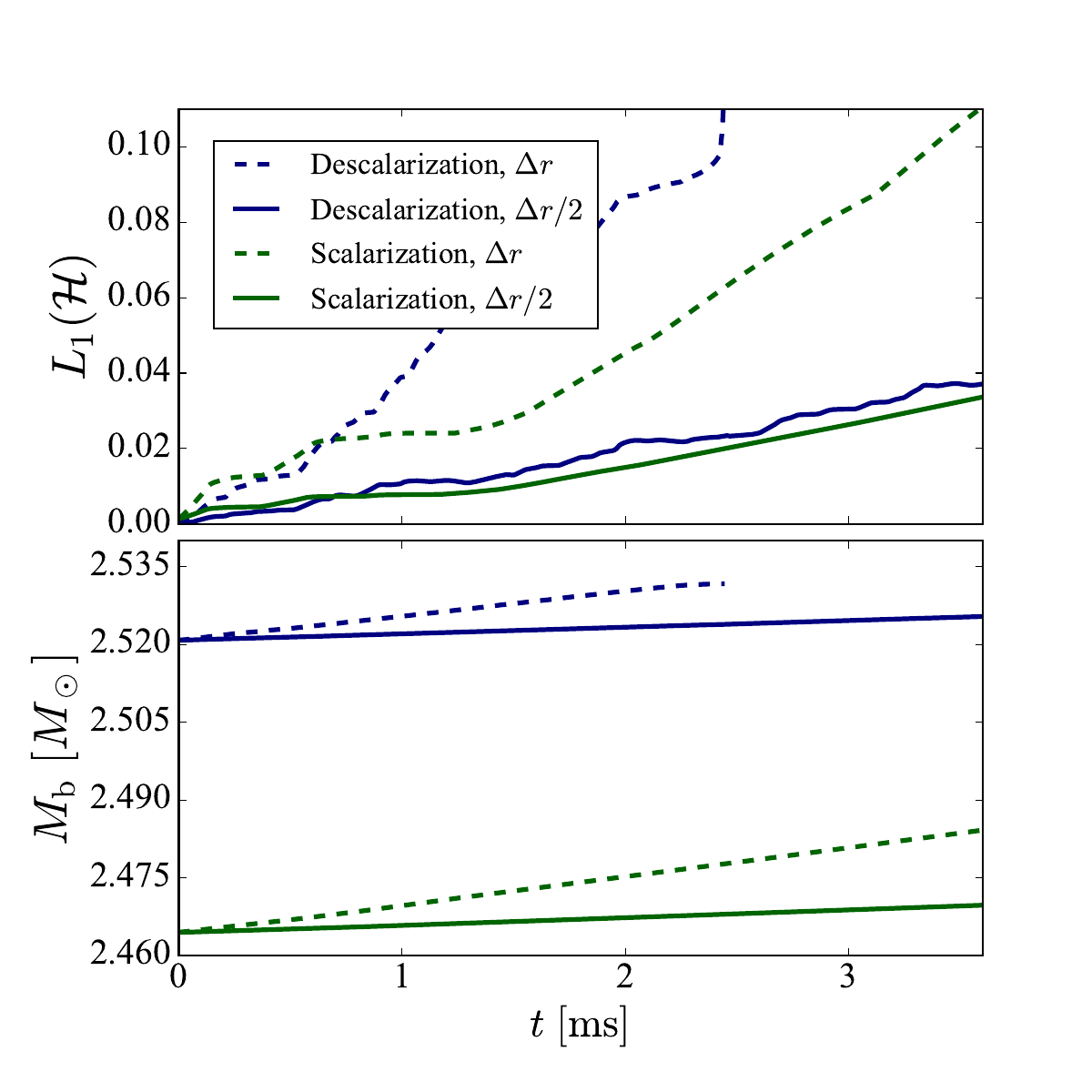}
     \caption{Time evolution of the $L_1$-norm of the Hamiltonian constraint ${\cal H}$ (upper panel) and the baryon mass $M_b$ (lower panel), for two consecutive numerical resolutions, during processes of scalarization and descalarization. At higher resolution, $M_b$ tends to remain constant in time, as expected. Correspondingly, $L_1({\cal H})$ decreases by a factor $\sim 3.5$.}
     \label{fig:H_and_Mb}
\end{figure}

In both cases, after applying the perturbation, we update the spacetime metric functions by solving the constraint equations, thereby generating consistent initial data for the time evolution. The upper panel of Fig.~\ref{fig:H_and_Mb} shows the $L_1$-norm of the Hamiltonian constraint violation as a function of time for two different numerical resolutions, in both the scalarization and descalarization scenarios. As expected, this residue gets strictly smaller as resolution increases. Our baseline numerical resolution corresponds to $\Delta r = 10^{-4} r_\text{max}$, where $r_\text{max} \sim 400$ km is the size of the radial domain.
The lower panel of Fig.~\ref{fig:H_and_Mb} shows the evolution of the baryon mass, which should ideally remain constant over time. Due to numerical errors, a slight increase $M_b$ is observed. However, we verify that this growth rate decreases with higher numerical resolution, and $M_b$ tends to stay constant over time.


\bibliography{lib}

\end{document}